\documentclass[%
 preprintnumbers,
 amsmath,amssymb,
 aps,
 floatfix,
]{revtex4-1}
\usepackage{hyperref}
\usepackage{amsmath}
\usepackage{amsfonts}
\usepackage{xspace}
\usepackage{graphicx}
\usepackage{xcolor}
\usepackage{dcolumn}           
\usepackage{bm}
\usepackage{mathtools}
\usepackage{algorithm}
\usepackage[noend]{algpseudocode}

\newcommand{\gras}[1]{\boldsymbol{#1}}

\newcommand{\pvec}{\boldsymbol{p}}

\newcommand{\setR}      {\mathbb{R}} 
\newcommand{\op}    [1] {\hat{#1}}   
\newcommand{\conC}  [1] {{#1}^\star} 

\newcommand{\conH}  [1] {{#1}^\dag}

\newcommand{\matx}  [2][e] 
{%
  \ifx#1e
    \mathrm{#2}
  \else\if#1h
    \conH{\mathrm{#2}}
  \else\if#1i
    \mathrm{#2}^{-1}
  \else\if#1t
    \trans{\mathrm{#2}}
  \else
    \mathrm{#2}
  \fi\fi\fi\fi
}%

\newcommand{\bra}   [1] {\ensuremath{\langle{#1}|}\xspace}
\newcommand{\ket}   [1] {\ensuremath{|{#1}\rangle}\xspace}
\newcommand{\braket}[2] {\ensuremath{\langle{#1}|{#2}\rangle}\xspace}

\newcommand{\projPN}  [3] {\ensuremath{\hat{P}_{#1}^{#2}(#3)}}
\newcommand{\countPN} [2] {\ensuremath{\hat{N}_{#1}^{#2}}}
\newcommand{\probPN}  [3] {\ensuremath{\mathbb{P}_{#1}^{#2}({#3})}}

\newcommand{\idPV}        {0}

\newcommand{\opCPw}   [2][] {\ensuremath{\op{c}^{#1\dag}({#2})}}

\newcommand{\opCPc}   [2][] {\op{c}^{#1\dag}_{#2}}
\newcommand{\opAPc}   [2][] {\op{c}^{#1}_{#2}}

\newcommand{\opCPr}   [2][] {\op{c}^{#1\dag}({#2})}
\newcommand{\opAPr}   [2][] {\op{c}^{#1}({#2})}

\newcommand{\idMFp}   [2][]
{%
  \ifx#1c
    \conC{\mathrm{\varphi}}_{#2}
  \else\if#1h
    \conH{\mathrm{\varphi}}_{#2}
  \else
    \mathrm{\varphi}_{#2}
  \fi\fi
}%

\newcommand{\trans} [1] {\ensuremath{{#1}^{\scriptscriptstyle\top}}}

\newcommand{\figref}[1]{Figure~\ref{#1}}
\newcommand{\tabref}[1]{Table~\ref{#1}}

\begin{document}

\preprint{LLNL-JRNL-760851}\preprint{LA-UR-18-30828}

\title{Number of Particles in Fission Fragments}
\date{August 14, 2019}

\author{Marc \surname{Verriere}}
\email[]{verriere@lanl.gov}
\affiliation{Los Alamos National Laboratory, Los Alamos, NM 87545, USA}

\author{Nicolas \surname{Schunck}}
\email[]{schunck1@llnl.gov}
\affiliation{Nuclear and Chemical Sciences Division, Lawrence Livermore National Laboratory, 
Livermore, CA 94551, USA}

\author{Toshihiko \surname{Kawano}}
\affiliation{Los Alamos National Laboratory, Los Alamos, NM 87545, USA}

\begin{abstract}
\begin{description}
\item[Background]
In current simulations of fission, the number of protons and neutrons in a given fission fragment 
is almost always obtained by integrating the total density of particles in the sector of space 
that contains the fragment. The semiclassical nature of this procedure and the antisymmetry of 
the many-body wave-function of the whole nucleus systematically leads to noninteger numbers of 
particles in the fragment. 
\item[Purpose]
We seek to estimate rigorously the probability of finding $Z$ protons and $N$ neutrons 
in a fission fragment, i.e., the dispersion in particle number (both charge and mass).  
Knowing the dispersion for any possible fragmentation of the fissioning nucleus will improve the 
accuracy of predictions of fission fragment distributions and the simulation of the fission spectrum with 
reaction models.
\item[Methods]
Given a partition of the full space $\mathbb{R}^3$ in two sectors corresponding to the two 
prefragments, we discuss two different methods. The first one is based on standard projection 
techniques extended to arbitrary partitions of space. We also introduce a novel sampling method 
that depends only on a relevant single-particle basis for the whole nucleus and the occupation 
probability of each basis function in each of the two sectors. We estimate the number of 
particles $A$ in the left (right) fragment by statistical sampling of the joint probability 
of having $A$ single-particle states in the left (right) sector of space.
\item[Results]
We use both methods to estimate the charge and mass number dispersion 
of several scission configurations in $^{240}$Pu using either a macroscopic-microscopic approach 
or full Hartree-Fock-Bogoliubov calculations. We show that restoring particle-number symmetry 
naturally produces odd-even effects in the charge probability, which could explain the well-known 
odd-even staggering effects of charge distributions.
\item[Conclusions]
We discuss two methods to estimate particle-number dispersion in fission fragments. In the limit 
of well-separated fragments, the two methods give identical results. It can then be advantageous 
to use the sampling method since it provides a $N$-body basis for each prefragment, which can be 
used to estimate fragment properties at scission. When the two fragments are still substantially 
entangled, the sampling method breaks down and one has to use projector techniques, which gives 
the exact particle-number dispersion even in that limit. Note that in this paper, we have assumed 
that scission configurations could be described well by a static Bogoliubov vacuum: the strong 
odd-even staggering in the charge distributions could be somewhat attenuated when going beyond 
this hypothesis.
\end{description}
\end{abstract}

\maketitle


\section{Introduction}
\label{sec:intro}

The theoretical understanding of nuclear fission, discovered in 1938 by O. Hahn and 
F. Strassmann, remains a vexing challenge even to this day. The fission of a heavy atomic 
nucleus presents a number of conceptual as well as practical difficulties. A fissioning 
nucleus is a particular example of a quantum many-body system of strongly interacting 
fermions, whose interaction is only known approximately. Fission dynamics is explicitly 
time dependent and involves open channels (mostly neutrons, but also photons). From a 
fundamental perspective, the physics of scission, or how an interacting, quantum many-body 
system splits into two well-separated interacting quantum many-body systems is very 
poorly known. Although there has been a considerable body of experimental work on fission 
in general, the most accurate data involve the decay of the fission fragments. The 
mechanism by which these fragments are formed in the first place must be described by theory.

Several approaches have been developed over the years to describe the fission process. Since 
fission times are rather slow compared with single-particle (s.p.) types of 
excitations~\cite{bulgac2016,bulgac2018}, quasistatic approaches are well justified. Most 
incarnations of these approaches rely on identifying a few collective variables that drive 
the fission process, mapping out the potential energy surface in this collective space (which 
fixes all properties of fission fragments) and computing the probability for the nucleus 
to be at any point on the surface, e.g., with semiclassical dynamics, such as Langevin~ 
\cite{abe1996,froebrich1998,usang2016}, random walk~\cite{randrup2011,randrup2011a,randrup2013,
ward2017} or with fully quantum-mechanical dynamics such as the time-dependent generator 
coordinate method~\cite{berger1984,goutte2005,regnier2016,regnier2018}. One major limitation 
of these approaches is the need to identify scission configurations in the potential energy 
surface, that is, the arbitrary frontier that separates configurations where the nucleus is 
whole from those where it has split into two fragments~\cite{younes2011,schunck2014,
schunck2016}. In practice, such scission configurations happen to always be characterized by 
noninteger values of average particle numbers in the fragments. 

The arbitrariness of the very concept of scission is strongly mitigated in explicitly 
nonadiabatic theories of fission, such as the various formulations of time-dependent nuclear 
density functional theory~\cite{umar2010,simenel2014,goddard2015,scamps2015,tanimura2015,
bulgac2016,goddard2016,tanimura2017,bulgac2018}. Since these approaches simulate the real-time 
evolution of the nucleus and explicitly conserve energy, one can obtain excellent estimates of 
fission fragment properties well past the actual scission point~\cite{simenel2014,bulgac2016,
bulgac2018}. However, these theories still simulate the evolution of the fissioning nucleus 
instead of the fragments themselves: The latter remain entangled even after scission and, thus, 
have also noninteger values of protons and neutrons~\cite{scamps2015}. Particle-number symmetry 
in the fragments could, in principle, be restored by using standard projection techniques. 
This approach was pioneered initially in the case of particle transfer in heavy-ion fusion 
reactions \cite{simenel2010,scamps2013,sekizawa2013,sekizawa2014}. However, as we will discuss 
below, projection techniques do not allow to identify independent configurations in the 
fragments. Although this does not impact the estimate of primary charge or mass yields, it would 
be very useful to have access to a basis of $N$-body Slater determinants of particles to 
calculate other observables.

Our goal is, thus, to explore an alternative method to estimate the number of particles in 
the fission fragments. More precisely, given a description of the fissioning system by an 
$A$-body Slater determinant or a quasiparticle vacuum, we seek to determine both the probability 
that the total $N$-body wave-function contains a Slater determinant of particles with $A_{1}$ 
($A_{2}$) particles in the left (right) fragment, both $A_1$ and $A_{2}$ being integers and 
$A_1 +  A_2 = A$, as well as a suitable basis of s.p. states to describe them.
In this paper, we propose a new method that only depends on a physically relevant s.p. 
basis for the fissioning nucleus and a set of occupation probabilities.

We present our theoretical framework in Section II. This includes some general notations, the 
presentation of our Monte Carlo (MC) sampling method, and a reminder of projection techniques 
adapted to the case of fission fragments. Section III is focused on the validation and 
numerical implementation of the sampling method. In Section IV, we study in more details the 
fragmentation probabilities for scission configurations in $^{240}$Pu, before concluding in 
Section V.


\section{Theoretical Formalism}
\label{sec:theory}

The prediction of the primary mass or charge distribution in fission can be decomposed in three 
steps \cite{schunck2016}. First, one calculates the energy of the fissioning system for a 
set of configurations with different geometric shapes or constraints. This potential energy 
surface (PES) is divided into two regions, one where the nucleus has split into two fragments, 
the other where it has not; scission configurations correspond to the frontier between the 
two regions. In a second step, one estimates the probability to populate each scission 
configuration. This can be done either semiclassically by solving the Langevin equation as in 
Refs.~\cite{nadtochy2005,nadtochy2007,sadhukhan2011,nadtochy2012,aritomo2015,sadhukhan2016,usang2016,
ishizuka2017,sierk2017,usang2017} or with a random walk approach \cite{randrup2011,randrup2011a,
randrup2013,moeller2015,ward2017}, or more microscopically by using time-dependent configuration 
techniques, such as the time-dependent generator coordinate method with the Gaussian overlap 
approximation (TDGCM-GOA)~\cite{goutte2004,goutte2005,regnier2016,zdeb2017,tao2017,zhao2019,
regnier2019}. Finally, the result of the time-dependent evolution are coupled with an estimate 
of fragment properties to estimate the actual distribution of nuclear observables, such as charge, 
mass, total kinetic energy (TKE), etc.

In the case of charge and mass distributions, the last step of the procedure outlined above 
involves estimating the particle number in the fission fragments. Until recently, all fission 
calculations have used a semiclassical estimate of the average particle number based on 
integrating the one-body density. Below, we describe two methods to obtain integer values of 
particle number. The first method, which is developed in this paper, involves a Monte Carlo 
sampling of s.p. configurations. Its main advantage is that 
it indirectly provides a set of s.p. wave-functions for each of the fragments, 
which could then be used to estimate quantities, such as, e.g., the level density. The second 
method is based on extending standard projection techniques to arbitrary space  
partitionings and was introduced originally by Simenel in \cite{simenel2010} for Slater 
determinants and later extended in \cite{scamps2013} to the case of superfluid systems. 


\subsection{Space partitioning}
\label{subsec:partition}

Let us first assume that the state $\ket{\Phi}$ of the fissioning system is a Slater 
determinant of particles,
\begin{equation}\label{eq:ideal:Nbodystate}
\ket{\Phi} \equiv \prod_{k=0}^{A-1} \opCPc{k}\ket{\idPV},
\end{equation}
where $\ket{\idPV}$ is the particle vacuum. The operator $\opCPc{k}$ creates a fermion in the 
s.p. state $k$ and reads
\begin{equation}
\opCPc{k} = \int_{\mathcal{V}}\mathrm{d}^3\gras{r}\,\varphi_k(\gras{r})\opCPw{\gras{r}},
\end{equation}
where $\varphi_k(\gras{r})$ is the s.p. wave-function for state $k$ and the 
operator $\opCPw{\gras{r}}$ creates a well-localized fermion at point $\gras{r}$ (we omit  
spin and isospin degrees of freedom for the sake of simplicity). Recall that the set of all 
functions $\varphi_k(\gras{r})$ forms a basis of the $\mathcal{L}_2$ Hilbert space of 
square-integrable functions.

\begin{figure}[!htb]
\includegraphics[scale=0.27]{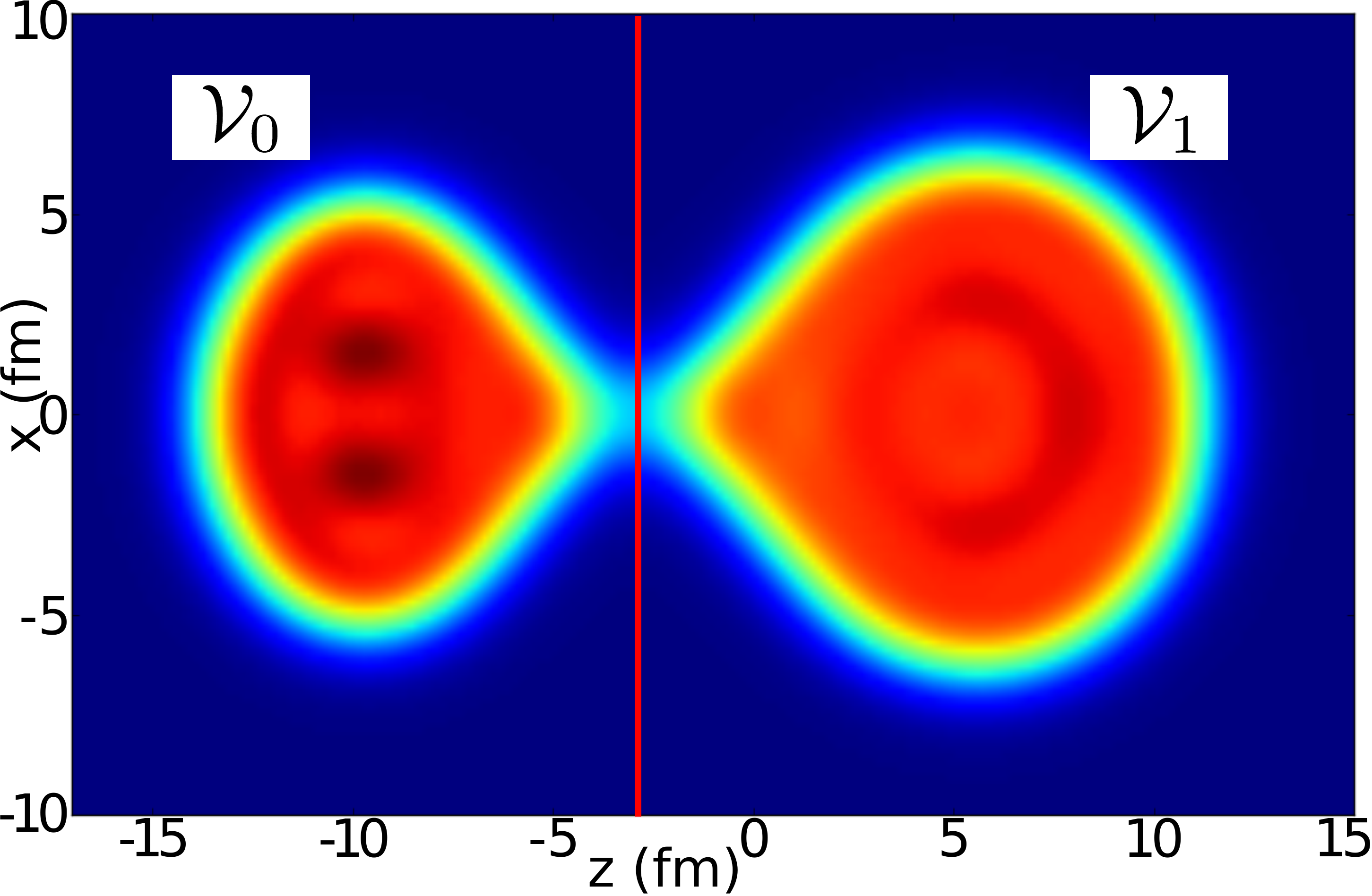}
\caption{\label{fig:ideal:denssplit}
Illustration of the partition of the space $\mathcal{V}$ into $\mathcal{V}_0$ and $\mathcal{V}_1$ 
using the local density. The color scale corresponds to the value of the local density associated 
with a Slater determinant $\ket{\Phi}$. The red line separates $\mathcal{V}_0$ (to the left of the 
line) and $\mathcal{V}_1$ (to the right of the line).}
\end{figure}

We also assume that it is possible to partition the full space $\mathcal{V} \equiv \setR^3$ 
into two sectors $\mathcal{V}_0$ and $\mathcal{V}_1$ such that $\mathcal{V}_0$ ($\mathcal{V}_1$) 
is the region where the left (right) fragment is localized. Such a partitioning could, for 
instance, be defined by introducing the coordinates of a neck between the two prefragments at 
scission as illustrated schematically in~\figref{fig:ideal:denssplit}. The two prefragments 
are separated by the red line located at the neck position. It is, then, always possible to 
decompose the s.p. wave-functions into
\begin{equation}\label{eq:ideal:fragsepbasefunc}
\varphi_{k}(\gras{r}) 
= \alpha_{k}^{(0)}\varphi^{(0)}_{k}(\gras{r}) 
+ \alpha_{k}^{(1)}\varphi^{(1)}_{k}(\gras{r}),
\end{equation}
where $\varphi^{(p)}_{k}(\gras{r})$ is defined in $\mathcal{V}_p$ ($p=0,1$) and 
$\alpha_{k}^{(p)}$ are normalization coefficients obtained by integrating the s.p.
wave-functions in the domain $\mathcal{V}_p$, 
\begin{equation}\label{eq:ideal:coeffs}
\alpha_{k}^{(p)} = \sqrt{\int_{\mathcal{V}_p} d^3\gras{r}\, |\varphi_k(\gras{r})|^2}.
\end{equation}
In terms of operators, the expansion of Eq.\eqref{eq:ideal:fragsepbasefunc} simply translates 
into
\begin{equation}\label{eq:ideal:fragsepbase}
\opCPc[]{k}
= \alpha_{k}^{(0)}\opCPc[(0)]{k}
+ \alpha_{k}^{(1)}\opCPc[(1)]{k}.
\end{equation}

The ladder operators $\opCPc[(0)]{k}$, $\opCPc[(1)]{k}$ and their Hermitian conjugates verify
the following anticommutation relations ($p,r = 0,1$)
\begin{align}
    \left\{\opCPc[(p)]{k}, \opCPc[(r)]{l}\right\} &= \left\{\opAPc[(p)]{k}, \opAPc[(r)]{l}\right\} = 0 \\
    \left\{\opCPc[(p)]{k}, \opAPc[(r)]{l}\right\} &=
        \delta_{pr}\int{\rm d}\gras{r}\ 
                      \varphi^{(p)}_{k}(\gras{r})\varphi^{(r)*}_{l}(\gras{r})
\end{align}
In the general case, $\{\opCPc[(p)]{k}, \opAPc[(p)]{l}\} \neq 0$ when $k\neq l$.
However, it is always possible to restore all the fermion anticommutation relations by 
orthonormalizing (for example with the Gram-Schmidt procedure) the bases $\{\opCPc[(0)]{k}\}$ 
and $\{\opCPc[(1)]{k}\}$ separately. Given these prerequisites, the goal of our method is to 
estimate the relative probability of finding a many-body state with $N_p$ particles in the 
subspace $\mathcal{V}_p$. 
 

\subsection{Monte Carlo Approach}
\label{subsec:MC}

We first present a method that only requires the coefficients \eqref{eq:ideal:coeffs}. 
Calculating them requires, in turn, only two ingredients: a set of s.p. wave-functions 
and a partitioning of $\setR^3$. Let us emphasize that from a mathematical or algorithmic point of 
view, the partitioning of the space is entirely arbitrary. We refer to our method as MC sampling.


\subsubsection{Orthonormal Bases}

We first introduce the general principles of our method for the idealized case where the 
$\varphi_{k}^{(p)}$ form an orthonormal basis of $\mathcal{V}_p$. We emphasize very clearly 
that, in the most general case, this condition is \emph{not} satisfied. In the context of fission, 
however, it can be approached asymptotically in the limit of infinitely separated fragments. 
In practice, it is reasonable to assume that scission configurations will sufficiently well 
approximate this limiting case so that the method can still provide reasonable estimates of 
the particle numbers. 

If the $\varphi_{k}^{(p)}$ form an orthonormal basis of $\mathcal{V}_p$, then the fermion 
anticommutation relations between the corresponding s.p. operators are satisfied. Let us 
insert Eq.~\eqref{eq:ideal:fragsepbase} in Eq.~\eqref{eq:ideal:Nbodystate}. We obtain
\begin{equation}\label{eq:ideal:decompo1}
\ket{\Phi} = \sum_{\substack{\pvec = (p_0,\dots ,p_{A-1})\\ p_k\in(0,1)}} 
\left[\prod_{k=0}^{A-1}\alpha_k^{(p_k)}\right]
\left[\prod_{k=0}^{A-1}\opCPc[(p_k)]{k}\right] \ket{\idPV}.
\end{equation}
In Eq.~\eqref{eq:ideal:decompo1}, we sum over all possible $A$-uplets of $0$ and $1$. Since 
we assume that the $\opCPc[(0)]{k}$ and $\opCPc[(1)]{k}$ correspond to orthonormal bases, the 
$A$-body state $\ket{\pvec}$ defined by
\begin{equation}
\ket{\pvec} = \prod_{k=0}^{A-1}\opCPc[(p_k)]{k}\ket{0}
\end{equation}
is a Slater determinant. By using the fermion anticommutation relations of the $\opCPc[(p)]{k}$,
we see that the set of all the possible $\ket{\pvec}$ forms an orthonormal basis of the $A$-body 
space. By construction, each state $\ket{\pvec}$ is an eigenvector of the particle-number operators 
for \emph{both} $\mathcal{V}_{0}$ and $\mathcal{V}_{1}$ and contains two sets of particles. The first 
set is completely in $\mathcal{V}_0$ and will contribute only to the left fragment; the second 
set is completely in $\mathcal{V}_1$ and will contribute only to the right fragment. Therefore, 
we can easily calculate the number of particles in the left (right) fragment for $\ket{\pvec}$ and 
$N_0(\pvec)$ ($N_1(\pvec)$). Since $p_k$ is either 0 or 1, it is easy to show that
\begin{equation}
N_0(\pvec) = \sum_{k=0}^{A-1} (1-p_k), \qquad
N_1(\pvec) = \sum_{k=0}^{A-1} p_k.
\end{equation}
and that $N_0(\pvec) + N_1(\pvec) = A$ as expected. We can, therefore, rewrite Eq.~\eqref{eq:ideal:decompo1} 
in the form
\begin{equation}
\ket{\Phi} = \sum_{N_0=0}^{A-1} c_{N_0}\ket{N_0},
\end{equation}
where $\ket{N_0}$ is the normalized component of $\ket{\Phi}$ with $N_0$ fermions in the left 
fragment, which is given by
\begin{align}
\ket{N_0} &\equiv \frac{1}{c_{N_0}}\sum_{\substack{\pvec = (p_0,\dots ,p_{A-1})\\N_0(p)=N_0}}
                      \left[\prod_{k=0}^{A-1}\alpha_k^{(p_k)}\right]\ket{p} , \\
c_{N_0} &\equiv \sqrt{
                    \sum_{\substack{\pvec = (p_0,\dots ,p_{A-1})\\N_0(p)=N_0}} 
                        \left(\prod_{k=0}^{A-1}\alpha_k^{(p_k)}\right)^2
                     }. \label{eq:ideal:probcalc}
\end{align}
Let us consider two different states $\ket{N_0}$ and $\ket{N_0'}$ such that $N_0\neq N_0'$. 
The states $\ket{N_0}$ and $\ket{N_0'}$ are expanded on disjointed subsets of the basis 
$\ket{\pvec}$. Since we already showed that this basis is orthonormal, it implies that the states 
$\ket{N_0}$ and $\ket{N_0'}$ are orthogonal and the squared norm of $\ket{N_0}$ is given by
\begin{equation}
\braket{N_0}{N_0} = 1.
\end{equation}
We can now define the probability $\probPN{0}{}{N_0}$ to measure the left fragment with 
$N_0$ particles as
\begin{equation}\label{eq:ideal:probdef}
\probPN{0}{}{N_0} 
\equiv \braket{\Phi}{N_0}\braket{N_0}{\Phi} 
= c_{N_0}^2.
\end{equation}
Calculating all the probabilities $\probPN{0}{}{N_0}$ using~\eqref{eq:ideal:probcalc} 
and~\eqref{eq:ideal:probdef} scales like $A\times 2^A$. Although this can certainly be done for 
nuclei with $A<30$, it becomes problematic in heavy systems, such as actinides. Instead, we can 
use a statistical approach to sample this probability. Specifically, we will use Monte Carlo 
sampling techniques to estimate the distribution of probability. For an $A$-body Slater 
determinant, this only requires drawing $A$ uniformly-distributed random numbers at each 
iteration.


\subsubsection{Non-orthonormal Bases}

As briefly mentioned earlier, the set of s.p. functions $\varphi_k^{(p)}(\gras{r})$ 
does not, in general, forms a basis of the subspace $\mathcal{V}_p$. Note that $\mathcal{V}_p$ 
is a Hilbert space very similar to the usual Hilbert space of square-integrable functions 
$\mathcal{L}^{2}(\mathbb{R}^3)$. Therefore, it could, in principle, be equipped with a proper 
basis. The problem is that such bases are not necessarily related to the original basis of 
functions $\varphi_k(\gras{r})$ through a simple relation, such as 
Eq.\eqref{eq:ideal:fragsepbasefunc}. 

The only case where the functions entering Eq.~\eqref{eq:ideal:fragsepbasefunc} do form a basis 
of their respective Hilbert space is when the two fragments are infinitely separated. This can 
be most easily seen from exactly solvable models. In one dimension, for example, a double 
harmonic-oscillator potential of the type $V(x) = \tfrac{\omega^{2}}{8a^2}(x-a)^2(x+a)^2$, 
somewhat simulates the potential well between two (identical) prefragments separated by an 
average distance of $2a$. At the limit of infinite separation ($a \rightarrow \infty$), the 
two harmonic oscillators completely decouple and the solution of the Schr\"odinger equation 
for the full system tends toward the sum of two harmonic oscillators shifted by $\pm a$; see, 
e.g., Ref~\cite{kleinert2009} for a comprehensive presentation. Note that a full treatment of the 
problem with path integrals would still lead to a nonzero tunneling probability between the 
two systems, which is beyond the scope of this paper. 

The point of this short discussion is that our hypothesis that the two sets of functions 
$\varphi_{k}^{(0)}$ and $\varphi_{k}^{(1)}$ are {\em approximately} orthonormal should be 
reasonable.


\subsubsection{Inclusion of Pairing Correlations}
\label{subsubsec:pairing}

Pairing correlations play an essential role in the fission process~\cite{simenel2014,
tanimura2015,bulgac2016}. In static calculations, they are typically described within the BCS or 
Hartree-Fock-Bogoliubov (HFB) approximations (with or without projection). In both cases, one can always define a set of 
s.p. wave-functions $\varphi_{k}(\gras{r})$ associated with the operators $\opCPc{k}$. 
This basis can be, for instance, made of the eigenstates of some realistic average potential 
(macroscopic-microscopic approaches) or of the nuclear mean field (Hartree-Fock theory), or it 
can be the canonical basis in the HFB theory. Together with s.p. states, pairing 
theories also provide the occupation amplitudes $u_k$ and $v_k$, such that $u_k^2+v_k^2 = 1$.

Based on these remarks, one can extend our method of calculation for the probability 
$\probPN{}{}{N_0}$ of finding $N_0$ particles in the left fragment in the presence of pairing 
correlations by performing two consecutive statistical samplings. We first draw random sets 
of $A$-occupied levels from the canonical basis based on the values of the probability 
amplitudes $u_k^2$ and $v_k^2$. For any such sample, we can then apply the method outlined 
in the previous sections. In more detail, the procedure is the following:
\begin{enumerate}
\item For each energy-level $k$ in the canonical basis, draw a uniformly distributed random 
number $0\leq r_k \leq 1$ and select the level for occupation if $r_k \leq v_{k}^2$. The 
Slater determinant $\ket{\tilde{\Phi}_{\gras{p}}}$, thus, formed out of all the occupied levels, 
occurs with the probability
\begin{equation}
\probPN{}{}{\gras{p}} = \prod_{\substack{o\\p_o=1}}v_o^2\prod_{\substack{n\\p_n=0}}u_n^2.
\end{equation}
\item For each such state $\ket{\tilde{\Phi}_{\gras{p}}}$ with good particle number, we 
calculate the probability $\probPN{}{}{N_0}$ that the left fragment has $N_0$ particles by 
using the method presented earlier.
\item We repeat this two-step sampling as many times as needed for the final probability 
distributions $\probPN{}{}{N_0}$ to converge. In practice, this requires on the order of a few 
thousands of iterations. 
\end{enumerate}
It is important to realize that the first step of the procedure described above can be used to 
estimate the probability $\probPN{}{}{N_0}$ that an arbitrary BCS or HFB state contains exactly 
$N_0$ particles. Therefore, it is an alternative way to project on particle number without 
introducing any projector. We will take advantage of this observation to validate our method.


\subsection{Projectors Method}
\label{subsec:projector}

The number of particles in fission fragments can also be recovered by using projector 
techniques as developed in Refs.~\cite{simenel2010,scamps2013}. Here, we give a brief summary of 
projection with emphasis on practical aspects and possible differences between the MC sampling 
presented previously. 

In the standard approach to projection, see e.g., Refs.~\cite{ring2004,schunck2019}, the particle-number 
projection (PNP) operator $\projPN{}{q}{N}$ reads
\begin{equation}
\projPN{}{q}{N} \equiv \frac{1}{2\pi}\int_{0}^{2\pi}{\rm d}\theta\, e^{i\theta(\countPN{}{q} - N)},
\end{equation}
where $q=\text{neutron,proton}$ is the type of particle and $\countPN{}{q}$ is the particle-number operator. The 
idea in~\cite{simenel2010} is the following: instead of using the particle number on the full 
space, one may define an operator $\countPN{p}{q}$ that count the number of particles only in the 
partition $\mathcal{V}_{p}$. This operator can be written
\begin{equation}\label{eq:NdecompoVi}
\countPN{p}{q} 
\equiv 
\int dx\, \opCPr{x}\opAPr{x}\Theta_{p}(\gras{r}),
\end{equation}
where, as usual, we note $x \equiv (\gras{r},\sigma, q)$ and $\int dx \equiv \sum_{\sigma} 
\int d^{3}\gras{r}$. In \eqref{eq:NdecompoVi}, $p$ indexes the partitions $\mathcal{V}_p$ and 
$\Theta_{p}(\gras{r})$ represents the indicator function of $\mathcal{V}_p$, i.e.,
\begin{equation}
    \Theta_p(\gras{r}) \equiv
      \begin{cases}
        1 & \text{if $\gras{r} \in \mathcal{V}_p$,} \\
        0 & \text{otherwise.}
      \end{cases}
\end{equation}
Obviously, we have the operatorial equality,
\begin{equation}
\sum_p \countPN{p}{q} = \countPN{}{q}.
\end{equation}
In Ref.~\cite{simenel2010}, $\mathcal{V}_{p}$ is denoted as $R$ and corresponds to the set of points with 
a positive value of $x$. In this case, $\Theta_{R}(\gras{r})\equiv\Theta(x)$ and, therefore, we obtain 
the equation (1) of Ref.~\cite{simenel2010}. Based on this definition, we can construct $\projPN{p}{q}{N}$, 
the particle-number projector on the partition $\mathcal{V}_p$, as follows:
\begin{equation}
\projPN{p}{q}{N} 
\equiv 
\frac{1}{2\pi}\int_{0}^{2\pi}{\rm d}\theta\, e^{i\theta(\countPN{p}{q} - N)}.
\end{equation}

Computing the action of $\projPN{p}{q}{N} $ on an arbitrary HFB vacuum $\ket{\Phi}$ can be done 
by defining shift operators $\hat{z}_{p} \equiv z^{\countPN{p}{q}}$ with $z = e^{i\theta}$ 
and following the same approach as in Ref.~\cite{dobaczewski2007}. By working directly with these field 
operators in coordinate space, one may show that 
\begin{equation}
\big( \countPN{p}{q} \big)^{n}\opCPr{x} = \opCPr{x}\big[ \Theta_{p}(x) + \countPN{p}{q}\big]^{n} ,
\end{equation}
which leads to 
\begin{equation}
\hat{z}_{p} \opCPr{x}\hat{z}_{p}^{-1} = e^{i\theta\Theta_{p}(x)} \opCPr{x} .
\end{equation}
By introducing the expansion of the field operators $\opCPr{x}$ on a basis, we arrive at
\begin{equation}
b^{\dagger}_{k}(\theta) \equiv \hat{z}_{p} \opCPc{k} \hat{z}_{p}^{-1} = \sum_{l} F_{kl}(\theta) \opCPc{l} ,
\label{eq:zR_ck}
\end{equation}
with
\begin{equation}
F_{kl}(\theta) 
= \int dx\, \varphi^{*}_{k}(x)\varphi_{l}(x)e^{i\theta\Theta_{p}(x)} 
= \delta_{kl} + S_{kl}^{(p)}(e^{i\theta} - 1),
\label{eq:Fmatrix}
\end{equation}
where $S_{kl}^{(p)}$ is the overlap matrix of the basis $\varphi_{k}^{(p)}$. We then find that the
action of the shift operator on the HFB vacuum reads, in the canonical basis,
\begin{equation}
\hat{z}_{p}\ket{\Phi} = \prod_{k>0} \Big[ u_{k} + v_{k}b_{k}^{\dagger}(\theta)b_{\bar{k}}^{\dagger}(\theta)\Big] \ket{0} .
\end{equation}
The calculation of the norm overlap can, then, be efficiently computed with the Pfaffian 
techniques of Ref.~\cite{robledo2009} as outlined in Ref.~\cite{scamps2015}.

In order to estimate the probability $\probPN{p}{q}{N}$ to find $N$ nucleons of type $q$ in the 
spatial partition $\mathcal{V}_p$, we must also make sure that the total number of particles in 
the fissioning system is restored. Therefore, the probability $\probPN{p}{q}{N}$ involves a
double-projection,
\begin{equation}
\probPN{p}{q}{N} \equiv \frac{\bra{\Phi}\projPN{p}{q}{N}\projPN{}{q}{N_{\rm tot}}\ket{\Phi}}{\bra{\Phi}\projPN{}{q}{N_{\rm tot}}\ket{\Phi}} ,
\end{equation}
where $N_{\rm tot}$ is the total number of particles of species $q$ for the fissioning nucleus. 
As mentioned in Ref.~\cite{scamps2013}, the double-projection involves the integration over two 
gauge angles $\theta$ and $\theta'$. One can easily show that the rotated-wave function will 
simply read
\begin{equation}
\hat{z}_{p}\hat{z}\ket{\Phi} 
= \prod_{k>0} \Big[ u_{k} + v_{k}b_{k}^{\dagger}(\theta,\theta')b_{\bar{k}}^{\dagger}(\theta,\theta')\Big] \ket{0} ,
\end{equation}
where $b^{\dagger}_{k}(\theta,\theta')$ is given by \eqref{eq:zR_ck}, only replacing 
$F_{kl}(\theta)$ by $F_{kl}(\theta,\theta') = F_{kl}(\theta)e^{i\theta'}$.


\section{Proofs of Principle}

In this section, we study how the MC sampling method can be used to restore particle number. 
First, we validate the approach against the projector method in a standard case of 
particle-number restoration for a fully-paired HFB vacuum. We then apply MC sampling to 
estimate fragmentation probabilities in different subspaces and analyze the numerical 
convergence of the method.

\subsection{Validation}
The validation consists of using our sampling method to compute the coefficients $c_N$ of the 
expansion of an arbitrary HFB state $|\Phi\rangle$ on good-particle-number Slater determinants,
\begin{equation}\label{eq:HFBexp}
\ket{\Phi} = \sum_{N} c_{N}\ket{N},
\quad \ket{N} = \frac{\projPN{}{}{N}\ket{\Phi}}{\sqrt{\bra{\Phi}\projPN{}{}{N}\ket{\Phi}}} .
\end{equation}
This is done simply by following Step 1 of the procedure discussed in Section~\ref{subsubsec:pairing}. We chose (arbitrarily) the nucleus $Z=60$ and $N=70$ for the tests. 
We used the code \texttt{HFBTHO 3} \cite{perez2017} to solve the HFB equation for this nucleus in a 
deformed HO basis of 16 shells (oscillator length: $b_0 = 2.0$ fm, $\beta_2 = 0.2$). We took 
the SkM* parametrization of the Skyrme functional, a surface-volume pairing interaction with 
$V_{0n} = V_{0p} = -250$ MeV and an infinite quasiparticle cutoff. Note that it does not  
matter if these characteristics are realistic or not: They were chosen exclusively to make 
sure there was a substantial amount of pairing correlations for both protons and neutrons. 

\begin{table}[!htb]
\caption{Comparison of the sampling method and exact particle number-projection for the 
calculation of the coefficients of the expansion of Eq.~\eqref{eq:HFBexp} for both protons 
and neutrons; see the text for additional details.}
\label{tab:results:pnp}
\centering
\begin{ruledtabular}
\begin{tabular}{ccccccccc}
& \multicolumn{2}{c}{$|c_{N}|^2$} & \multicolumn{2}{c}{$|c_{Z}|^2$} \\
Number & PNP & sampling & PNP & sampling \\
\hline
N     & 0.370{\bf 6} & 0.370{\bf 7} & 0.390{\bf 6} & 0.390{\bf 5} \\
N-2   & 0.320{\bf 8} & 0.320{\bf 7} & 0.336{\bf 3} & 0.336{\bf 2} \\
N-4   & 0.2079       & 0.2079       & 0.197{\bf 1} & 0.197{\bf 2} \\
N-6   & 0.100{\bf 6} & 0.100{\bf 7} & 0.0760       & 0.0760 \\
\end{tabular}
\end{ruledtabular}
\end{table}

We then projected the HFB solution on $N_0 = 70$, $68$, $66$ and $64$ as well as on $Z_0 = 60$, 
$58$, $56$ and $54$ using the Fomenko discretization of the particle-number projector with 
$L = 13$ gauge points. Since number parity is conserved in this case, the coefficients $c_{N}$ 
of the expansion of Eq.~\eqref{eq:HFBexp} are then simply given by~\cite{anguiano2001,
sheikh2002,stoitsov2005}
\begin{equation} 
c_{N}^2 
= \langle \Phi|\projPN{}{}{N}|\Phi\rangle 
= \frac{1}{L} \sum_{l=0}^{L-1} \bra{\Phi} e^{i\theta_{l}(\hat{N} - N)} \ket{\Phi} ,
\end{equation}
where $\theta_{l} = \pi l/L$ are the gauge angles. To ensure that $\sum_{N\in I} |c_{N}|^2 = 1$ 
for our subset $I$ of particle numbers, we renormalized the coefficients. The table 
\ref{tab:results:pnp} compares the results obtained with direct projection and with our 
sampling method applied on the canonical basis. They are exact to within $10^{-4}$, which 
corresponds to the precision of the sampling.

\subsection{Fragmentation Probabilities}
We now examine how MC sampling can be used to estimate the particle number in different 
subspaces. For simplicity, we focused on the nucleus \textsuperscript{240}Pu. For this 
proof of principle, we consider a macroscopic-microscopic approach where the shape of the 
nucleus is described by the matched quadratic surface (3QS) parametrization~\cite{nix1969,
nix1972,hasse1988}. The s.p. states are obtained by solving the Schr\"odinger 
equation for a few specific elongated shapes listed in \tabref{tab:results:asym_shape} in 
an axial HO basis of $N_{\rm sh}=35$ shells. Pairing correlations are 
treated in the Lipkin-Nogami approximation with a seniority pairing force characterized 
by Eq. (107) of~\cite{moeller2016}. We used an energy window of $\pm 5$ MeV around the 
Fermi level to define our valence space; further details of the theoretical framework can 
be found in Ref.~\cite{moeller2016}. 

The quantities $A_{\rm L}$ and $Z_{\rm L}$ listed in \tabref{tab:results:asym_shape} refer to 
the mass and charge of the prefragments obtained by integrating the one-body local density at
the left and right of the neck position. The latter is defined as the point with the lowest
density between the two prefragments.
    
\begin{table}[!htb]
\caption{Characteristics of scission configurations: shape parameters, and average mass and charge 
fragmentation.}
\label{tab:results:asym_shape}
\centering
\begin{ruledtabular}
\begin{tabular}{ccrrrrrrr}
Shape & $\alpha_2$ & $\alpha_3$ & $\sigma_1$ & $\sigma_2$ & $\sigma_3$ & $A_{\rm L}$ & $Z_{\rm L}$ \\
\hline
I    & 0.30 & 0.192 & 3.500 & -0.576 & 0.640 & 99.61 & 39.78\\
II   & 0.25 & 0.203 & 3.889 & -0.365 & 0.810 & 101.33 & 40.91\\
III   & 0.25 & 0.250 & 3.500 & -0.450 & 1.000 & 102.36 & 41.82\\
IV   & 0.20 & 0.605 & 3.182 & -0.545 & 1.210 & 112.29 & 44.95\\
\end{tabular}
\end{ruledtabular}
\end{table}

We have calculated the fragmentation probabilities associated with the configurations of 
\tabref{tab:results:asym_shape}. We drew $n_{\rm pair}=10\,000$ $A$-body Slater determinants 
and for each of them used $n_{\rm loc}=10\,000$ Monte Carlo samples to estimate the number 
of particles in the fragments. Since the configurations are not fully scissioned, we take 
into account the uncertainty associated with the position $z_{\mathrm{neck}}$ of the neck by 
assuming that $z_{\mathrm{neck}}$ follows a normal distribution 
$\mathcal{N}(\bar{z}_{\mathrm{neck}}, A_{\mathrm{neck}} \times Q_{\mathrm{neck}})$, where 
$\bar{z}_{\mathrm{neck}}$ is the position of the minimum between the two fragments of the 
local density along the z-axis, $A_{\mathrm{neck}}=1$ fm/nuc and $Q_{\mathrm{neck}}$ is the 
average value of the Gaussian neck operator~\cite{berger1990,schunck2014}. 
    
\begin{figure}[!htb]
\centering
\includegraphics[scale=0.36]{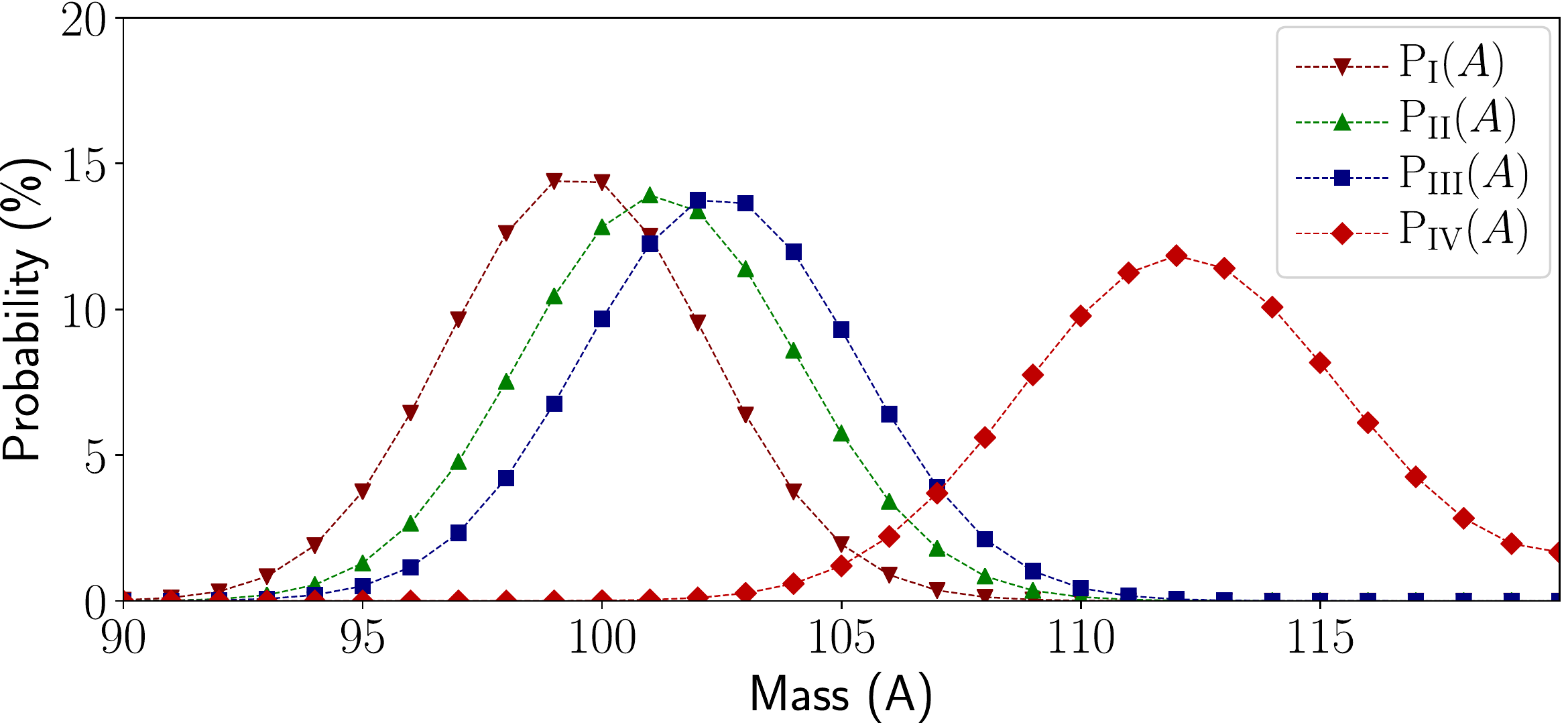}
\caption{Mass fragmentation probabilities (light fragment) for all the configurations listed in 
Table~\ref{tab:results:asym_shape}.}
\label{fig:res:YA_asym}
\end{figure}

\begin{figure}[!htb]
\centering
\includegraphics[scale=0.36]{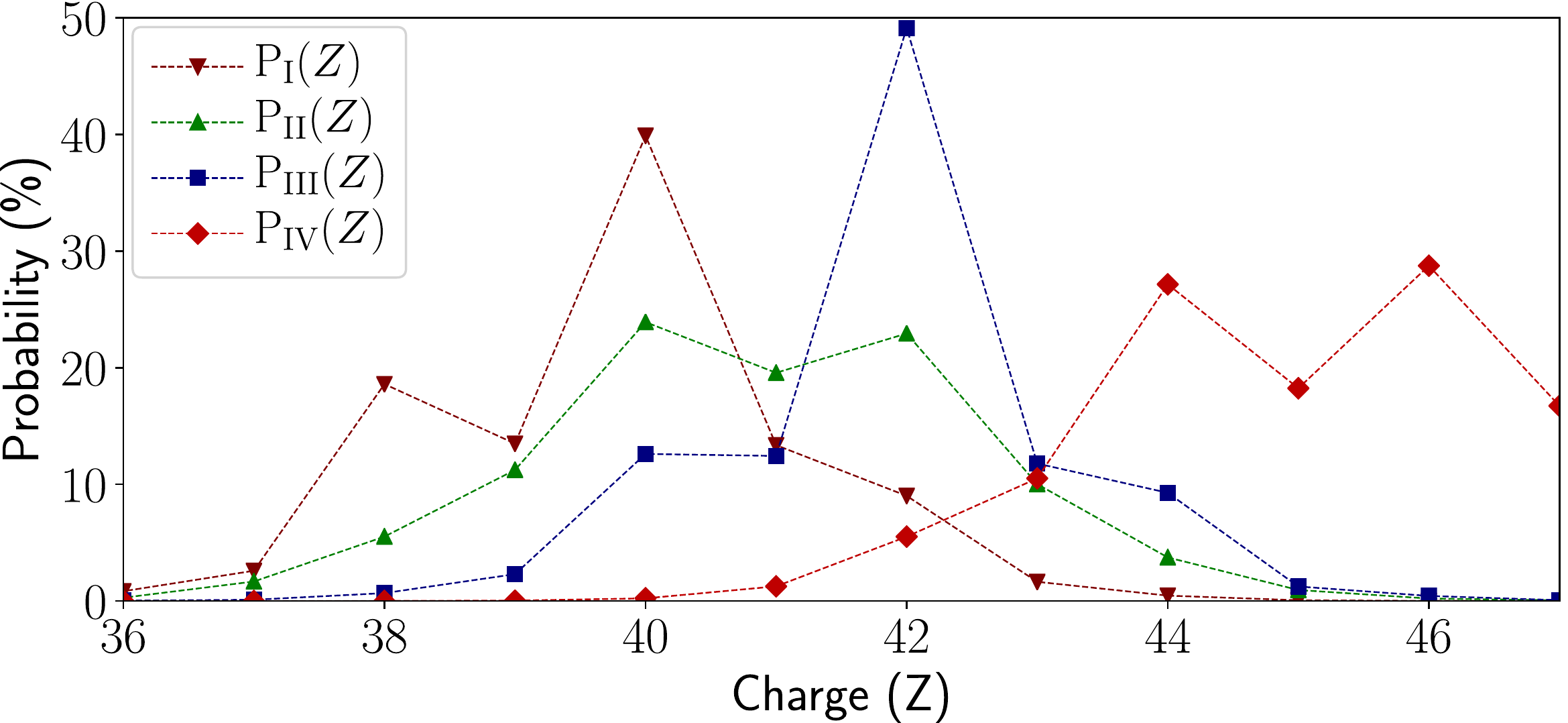}
\caption{Charge fragmentation probabilities (light fragment) for all the configurations listed in 
Table~\ref{tab:results:asym_shape}.}
\label{fig:res:YZ_asym}
\end{figure}

The mass fragmentation probabilities are shown in \figref{fig:res:YA_asym}. We note that 
all curves are smooth and are peaked near the values of $A$ corresponding to the geometric 
split between the fragments. There is no visible odd-even staggering (OES) for any of the mass 
probabilities. In the case of the charge fragmentation probabilities shown in 
\figref{fig:res:YZ_asym}, we note that the maximum of each curve is always associated with 
an even number of protons. Moreover, the probability for any even-proton number is always 
higher than the probability of any of the two odd-proton neighbors. In other words, we 
observe a clear OES. The different behavior of the mass and charge 
fragmentations is most likely caused by the Coulomb potential between the fragments, which 
increases the height of the effective barrier between them and, therefore, tends to localize 
the protons better than the neutrons. 

Note that the fragmentation probabilities shown in \figref{fig:res:YA_asym} and \figref{fig:res:YZ_asym} 
cannot be compared to experimental data~\cite{schillebeeckx1992,nishio1995}: They give 
only the dispersion around four specific fragmentations. In contrast, experimental fission fragment 
distributions include all possible fragmentations of the compound nucleus. To compare with 
experimental data, the first step is to explicitly simulate the nuclear dynamics, e.g., with 
semiclassical methods, such as Langevin or random walk \cite{nadtochy2005,nadtochy2007,sadhukhan2011,
nadtochy2012,aritomo2015,sadhukhan2016,usang2016,ishizuka2017,sierk2017,usang2017,randrup2011,
randrup2011a,randrup2013,moeller2015,ward2017} or microscopic methods, such as the time-dependent 
generator coordinate method \cite{goutte2004,goutte2005,regnier2016,zdeb2017,tao2017,zhao2019,
regnier2019}. This would provide the probability distribution $\probPN{}{}{\mathcal{S}}$ for the 
nucleus to be in a given scissionned or quasiscissionned state $\mathcal{S}$. The second step 
would be to fold the probability distribution thus obtained with the probabilities 
$\probPN{\mathcal{S}}{}{A}$ or $\probPN{\mathcal{S}}{}{Z}$ that our method provides via
\begin{align}
Y(X)   &= \sum_{\mathcal{S}} \probPN{}{}{\mathcal{S}} \probPN{\mathcal{S}}{}{X},\quad X=Z,A \\
Y(A,Z) &= \sum_{\mathcal{S}} \probPN{}{}{\mathcal{S}} \probPN{\mathcal{S}}{}{A}
\probPN{\mathcal{S}}{}{Z}.
\end{align}
Note that even if we do not consider correlations between protons and neutrons in the fragment 
probabilities in our method, the yields $Y(A,Z)$ obtained with the dynamics contains them.

\subsection{Numerical Convergence}\label{subsec:numconv}
The MC method of Sec.~\ref{subsec:MC} is statistical and relies on sampling a probability 
distribution. When pairing correlations are included, the sampling is characterized by two 
numbers: $n_{\rm pair}$, the number of draws of the particle Slater determinants (first 
step of the algorithm presented in Sec.~\ref{subsubsec:pairing}, p.~\pageref{subsubsec:pairing}), 
and $n_{\rm loc}$, the number of draws of a localized Slater determinant of $A$ particles 
from the left or right basis [cf. Eq.~\eqref{eq:ideal:decompo1}]. Here, we evaluate the 
uncertainty associated with these two integers for the particular case of configuration II 
in \tabref{tab:results:asym_shape}.

\begin{figure}[!htb]
\centering
\includegraphics[scale=0.35]{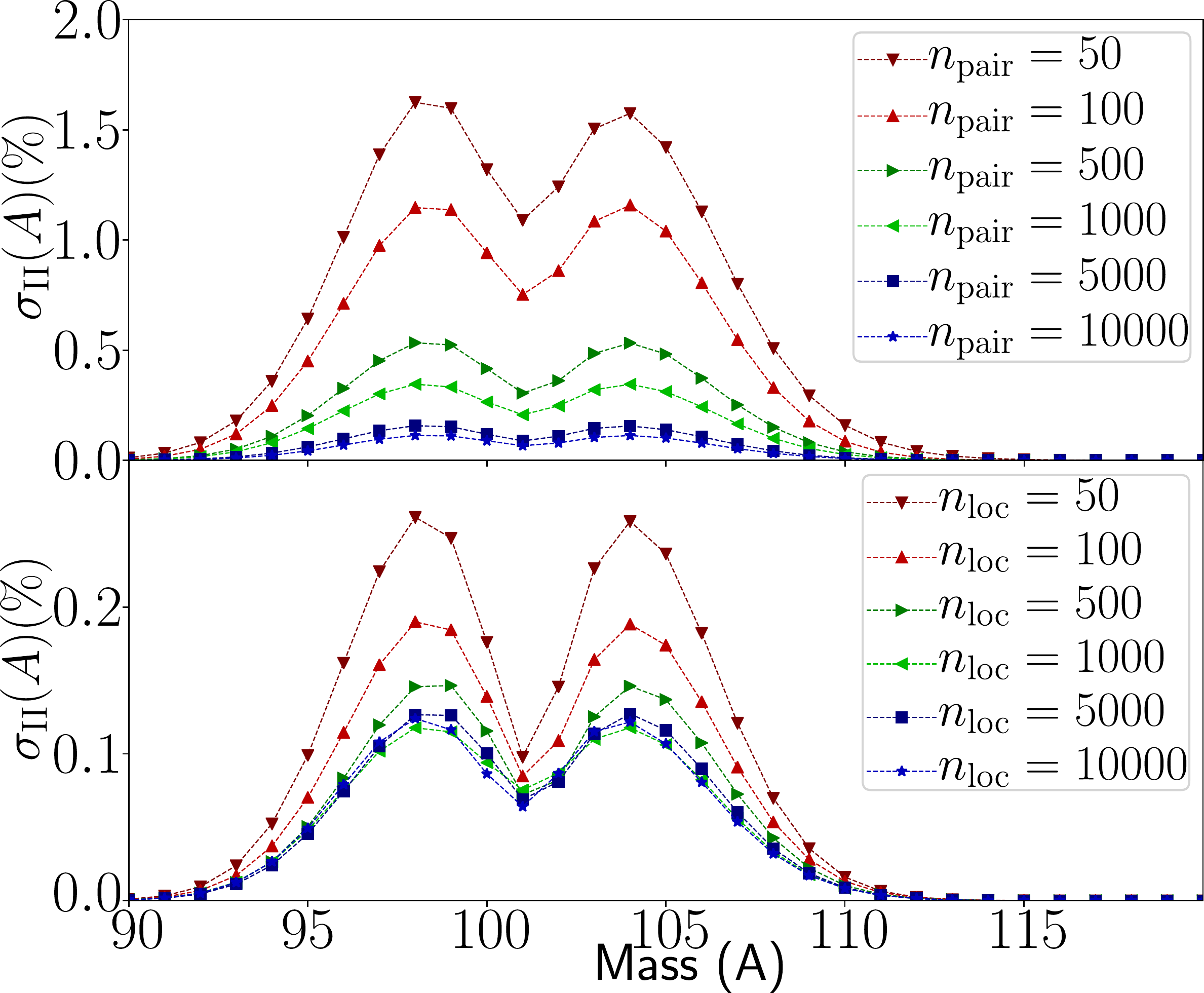}
\caption{Standard deviations associated with the probabilities to measure the light
fragment with a mass $A$ for the shape II. Top: For different values of $n_{\rm pair}$
and with $n_{\rm loc}=10\,000$. 
Bottom: For different values of $n_{\rm loc}$ and with $n_{\rm pair}=10\,000$.}
\label{fig:res:std_A}
\end{figure}

\begin{figure}[!htb]
\centering
\includegraphics[scale=0.35]{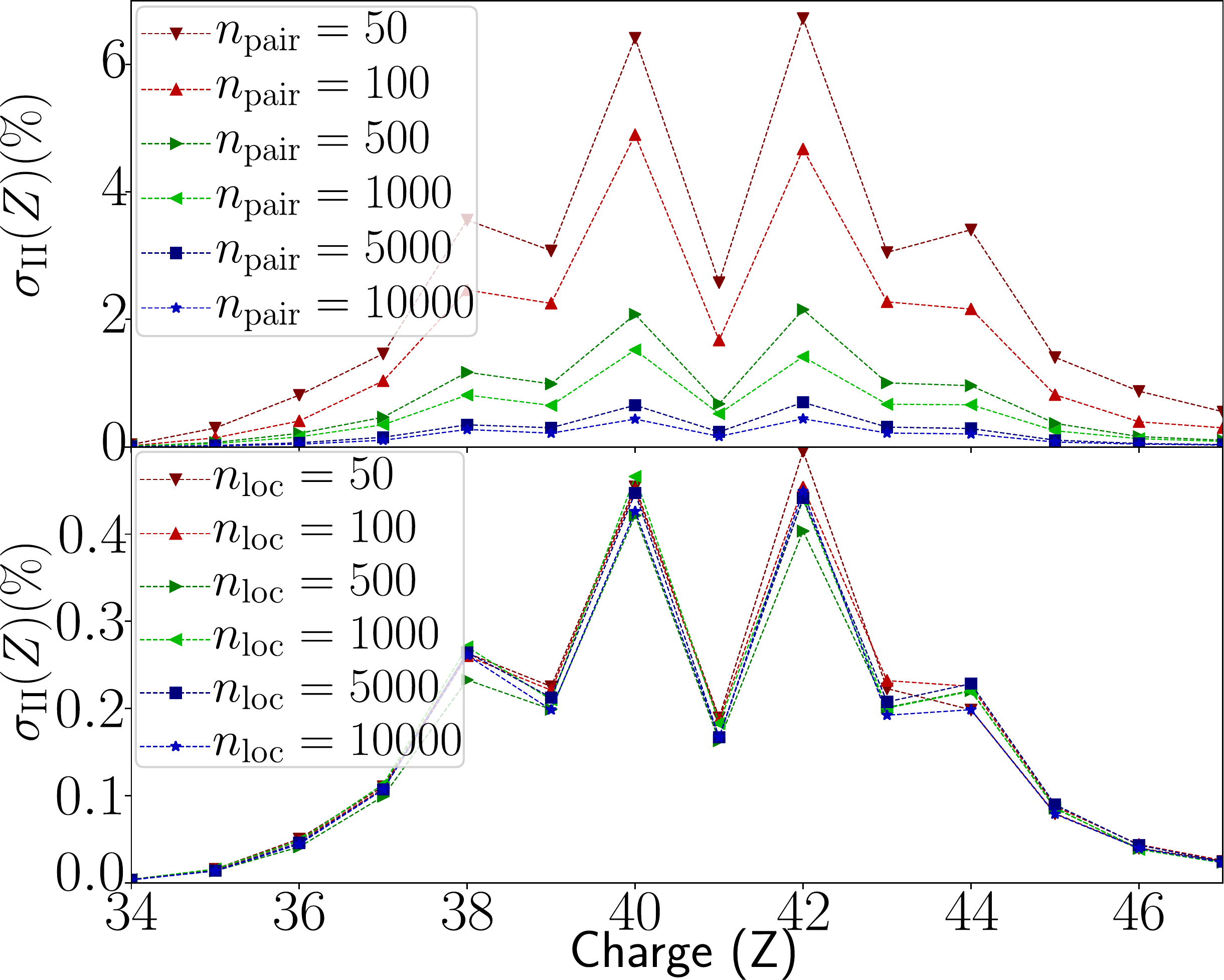}
\caption{Same as \figref{fig:res:std_A} for the charge fragmentations.}
\label{fig:res:std_Z}
\end{figure}
    
We considered 12 cases: $n_{\rm pair}=50$, $100$, $500$, $1\,000$, $5\,000$, $10\,000$ with
$n_{\rm loc}=10\,000$; and $n_{\rm pair}=10000$ with $n_{\rm loc}=50$, $100$, $500$, $1\,000$,
$5\,000$, $10\,000$. For each of them, we calculated the fragmentation probability in mass and
charge $S=200$ times. We then calculated the unbiased 
estimator of the standard deviation $\sigma_{\mathrm{II}}(X)$ for the distributions in mass 
and in charge using the following expression, for $X=A,Z$:
\begin{equation}
\sigma_{\mathrm{II}}(X)
\equiv
\sqrt{\frac{1}{S-1}
\sum_{k=0}^{S-1} \big[ Y^{(k)}_{\mathrm{II}}(X) - \bar{Y}_{\mathrm{II}}(X) \big]^2},
\end{equation}
where $Y^{(k)}_{\mathrm{II}}(X)$ is the $k$th calculation of the yields with our method and
$\bar{Y}_{\mathrm{II}}(X)$ is the mean value of all the $Y^{(k)}_{\mathrm{II}}(X)$. The 
standard deviations are shown in~\figref{fig:res:std_A} for the mass distributions, and 
in~\figref{fig:res:std_Z} for the charge distributions. The most sensitive parameter is 
$n_{\rm pair}$ with an improvement of 1.6\% of the standard deviations on the masses and 5.5\%
on the charges between the cases $n_{\rm pair}=50,n_{\rm loc}=10\,000$ and
$n_{\rm pair}=n_{\rm loc}=10\,000$ as shown in the upper panels of~\figref{fig:res:std_A}
and~\figref{fig:res:std_Z}. For all the cases with $n_{\rm pair}=10\,000$, the 
standard deviations are always below 0.5\%, and the improvement of the standard deviations is 
below 0.2\% for the masses and not visible for the charges between the cases $n_{\rm loc}=50$ and 
$n_{\rm loc}=10\,000$.
    
\begin{figure}[!htb]
\centering
\includegraphics[scale=0.36]{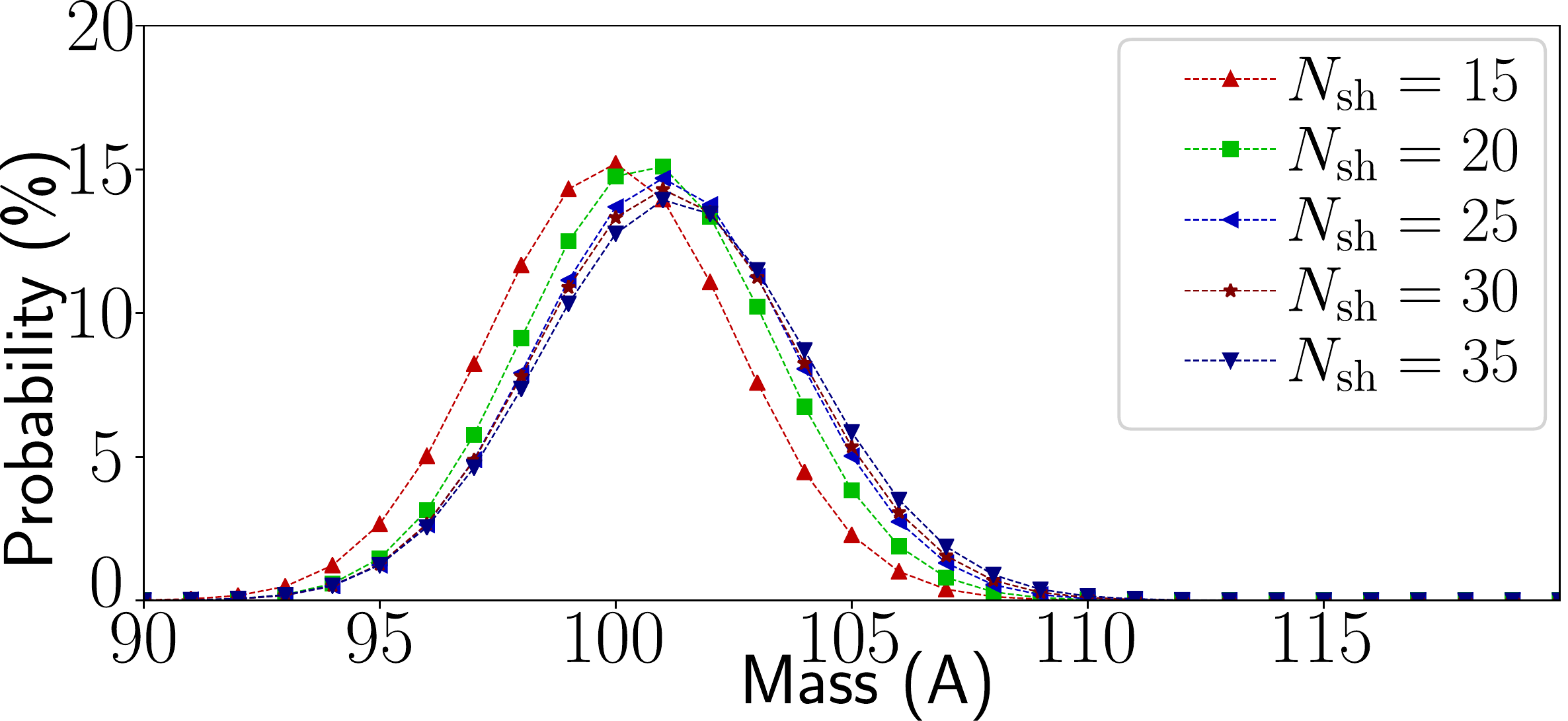}
\caption{Probabilities to measure the light fragment with a mass $A$ for the shape II
for different number of shells $N_{\mathrm{sh}}$ in the basis.}
\label{fig:res:sherr_A}
\end{figure}
    
\begin{figure}[!htb]
\centering
\includegraphics[scale=0.36]{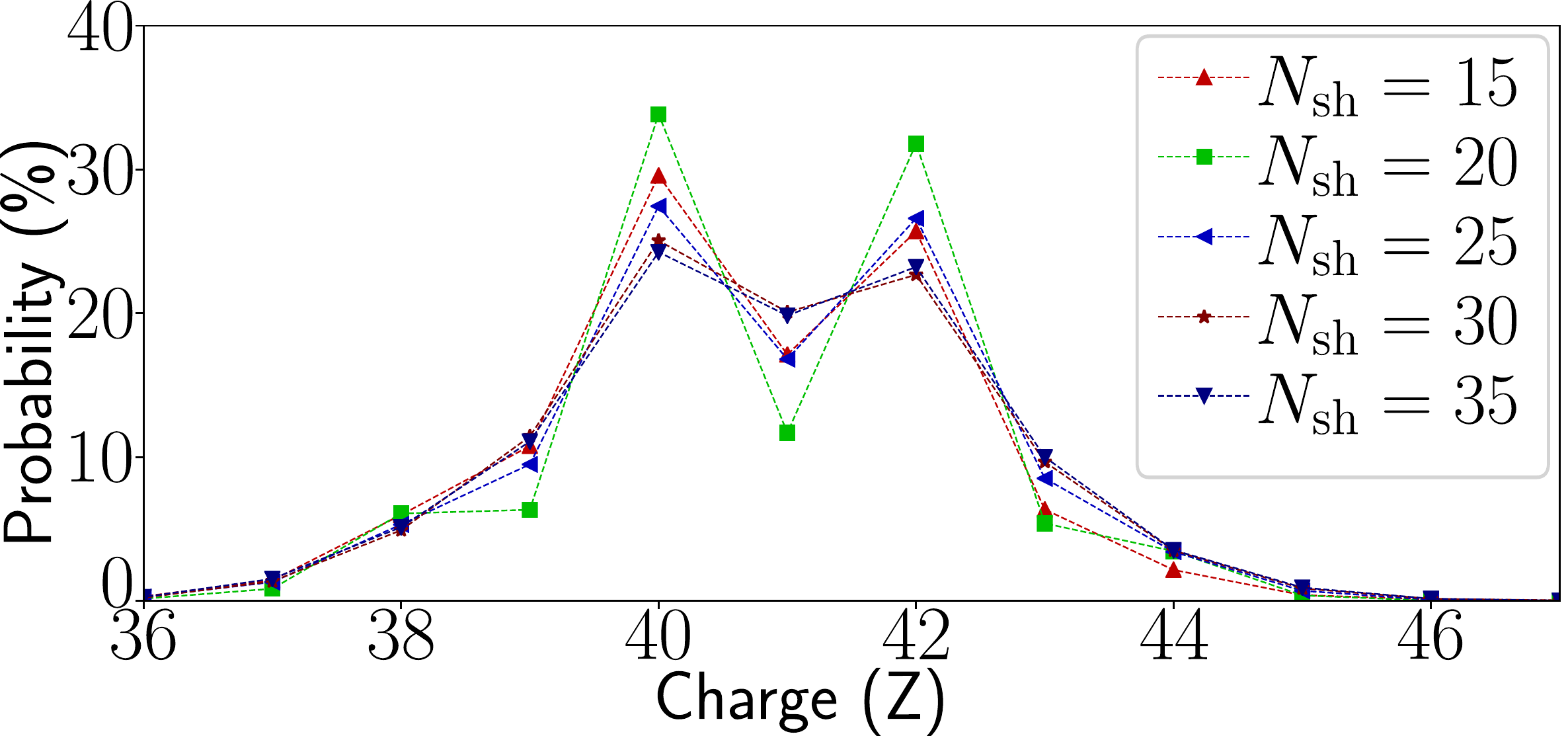}
\caption{Same as \figref{fig:res:sherr_A} for the charge of the light fragment.}
\label{fig:res:sherr_Z}
\end{figure}
    
The two other numerical parameters in our method are the number of shells $N_{\mathrm{sh}}$ 
and the truncation in the BCS occupation numbers $v^2_{\mathrm{thresh}}$. To analyze the 
impact of $N_{\mathrm{sh}}$, we have calculated the fragmentation probabilities in mass 
and charge for $N_{\mathrm{sh}} = 15, 20, 25, 30,$ and $35$ in the case of the configuration 
II. The corresponding curves are presented in~\figref{fig:res:sherr_A} 
and~\figref{fig:res:sherr_Z}. The increase in the number of shells shifts the most probable 
mass of the light fragment from $A=100$ to $A=101$ (and, therefore, shifts the most probable 
mass of the heavy fragment from $A=140$ to $A=139$). The increase of the number of shells 
does not change the most probable charge of the fragments. However, it drastically reduces 
the OES in the charge distribution between the cases $N_{\mathrm{sh}}=15$ 
and $N_{\mathrm{sh}}=35$. 
 
\begin{figure}[!htb]
\centering
\includegraphics[scale=0.36]{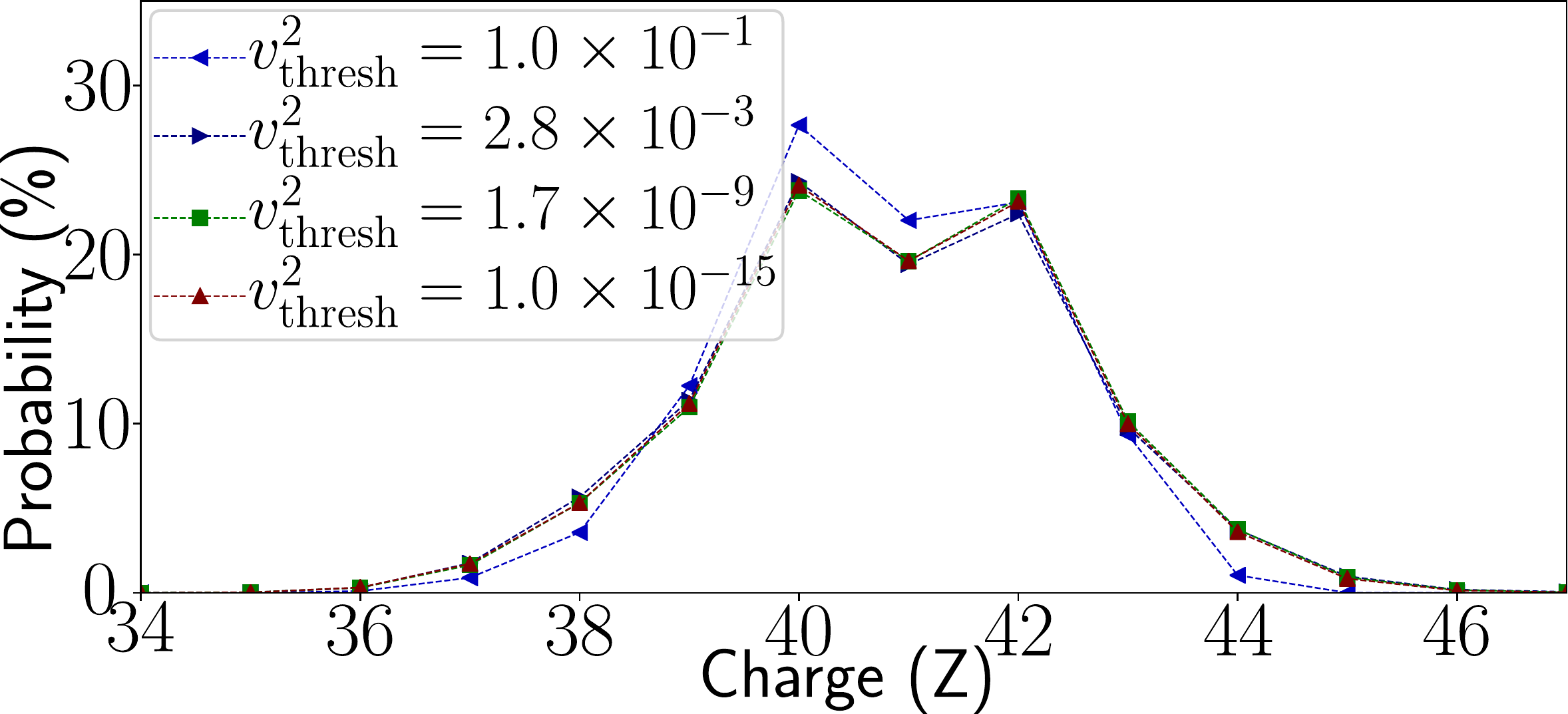}
\caption{Probabilities to measure the light fragment with a charge $Z$ for the shape II
different values $v^2_{\mathrm{thresh}}$ of the occupation numbers' threshold.}
\label{fig:res:verr_Z}
\end{figure}
    
To study the influence of the BCS occupations $v^2_{\mathrm{thresh}}$, we have calculated the 
fragmentation probability in mass and charge for the four cases $v^2_{\mathrm{thresh}}=1.0\times 
10^{-1}, 2.8\times 10^{-3}, 1.7\times 10^{-9}, 1.0\times 10^{-15}$. The corresponding 
probabilities are shown in~\figref{fig:res:verr_Z} for the charge distributions only. All the 
distributions for the mass and the charges have converged below 1\% for 
$v^2_{\mathrm{thresh}} \leq 2.8\times 10^{-3}$.


\section{Number of Particles at Scission}

In this section, we apply both the MC sampling and the projector methods to estimate the 
dispersion in particle number of realistic scission configurations for the case of 
\textsuperscript{240}Pu. 

\subsection{Macroscopic-microscopic Calculations}
First, we compare our approach with the projector method presented in Refs.~\cite{simenel2010,
scamps2013} for the macroscopic-microscopic approach. The Sch\"odinger equation was solved in 
a basis of $N_{\rm sh} = 35$ shells. Pairing correlations are treated in the same way as in 
Sec.~\ref{subsec:numconv}. We consider a scission configuration in \textsuperscript{240}Pu 
characterized by the following values of the 3QS parameters: $r_{\rm neck} = 2.50$ fm, 
$\alpha_2 = 0.448$, $\alpha_3 = 0.6259$, $\sigma_1 = 3.0613$, $\sigma_2 = -0.5349$, and 
$\sigma_3 = 0.9047$. This configuration corresponds to the most likely trajectory for a series 
of random walks on the five-dimensional potential energy surface~\cite{verriere2019}.

\begin{figure}[!htb]
\centering
\includegraphics[scale=0.25]{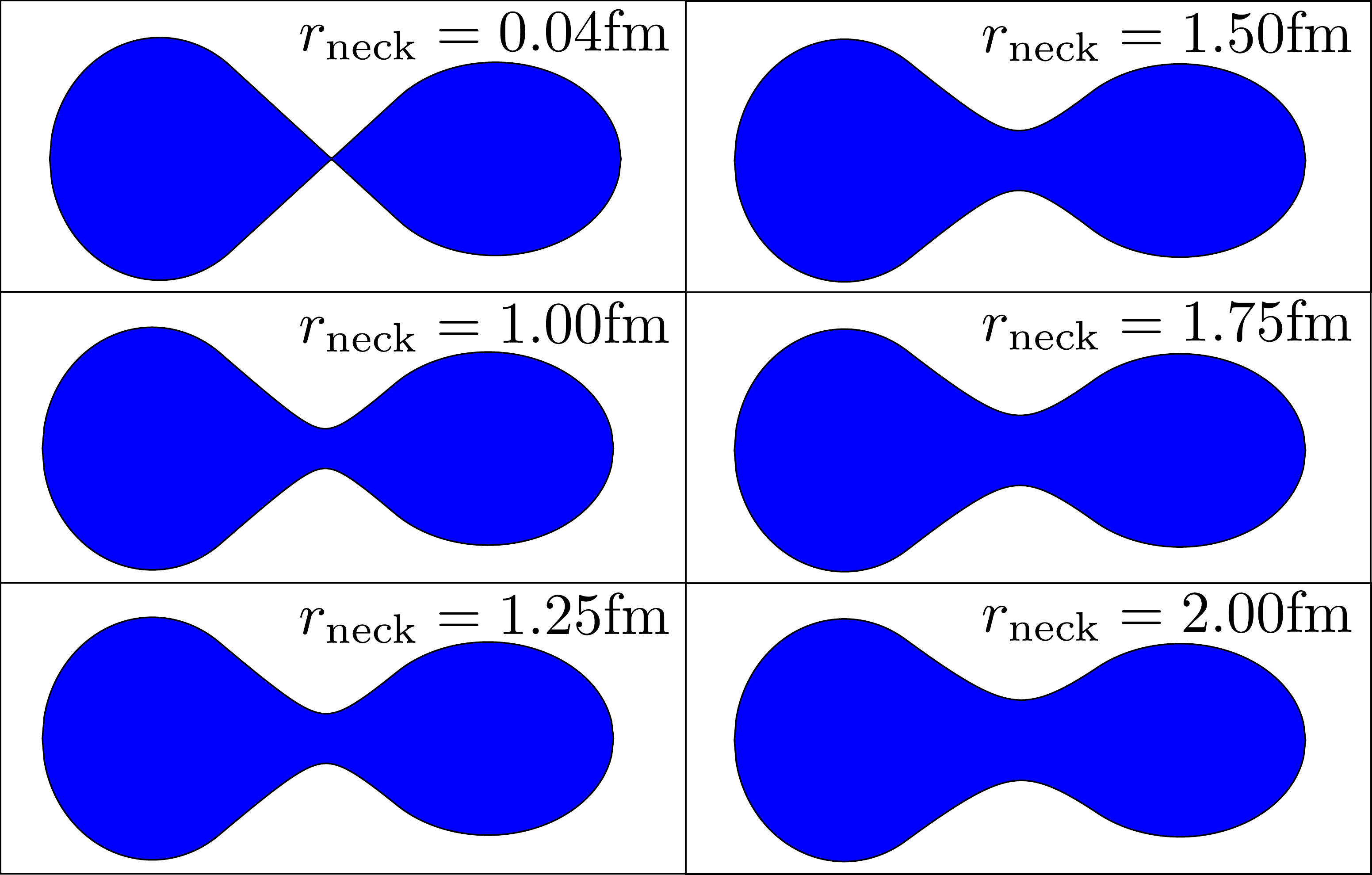}
\caption{Geometric shapes associated with the scission configurations of~\tabref{tab:results:neck_shape}.}
\label{fig:res:shapes_neck}
\end{figure}

Starting from this initial configuration, we vary the parameter $\sigma_2$ in order to reduce 
the size of the neck, and thus, approach the limit of two orthonormal s.p. bases 
for each of the prefragments. In practice, the values of $\sigma_2$ were chosen such that 
the neck radius takes the values $0.04$, $1.0$, $1.25$, $1.5$, $1.75$, $2.0$ fm. 
\figref{fig:res:shapes_neck} gives a visual representation of these configurations. For the 
MC sampling, we use $n_{\rm pair} = 10\,000$ and $n_{\rm loc} = 1\,000$; together with our basis
size of a $N_{\rm sh} = 30$ and a BCS threshold of $v^{2}_{\rm thresh} = 10^{-5}$, this gives a 
statistical precision of approximately 0.5\% on the charge and 0.15\% on the mass fragmentation 
probabilities.

\begin{table}[!htb]
\caption{Characteristics of scission configurations: neck radius, shape parameters and
average mass/charge fragmentations of the left fragment.}
\label{tab:results:neck_shape}
\centering
\begin{ruledtabular}
\begin{tabular}{cccccccc}
$r_{\rm neck}$ & $\alpha_2$ & $\alpha_3$ & $\sigma_1$ & $\sigma_2$ & $\sigma_3$ & $A_{\rm L}$ & $Z_{\rm L}$ \\
\hline
0.04 & 0.448 & 0.6259 & 3.0613 & -0.8431 & 0.9047 & 138.8 & 53.1 \\
1.00 & 0.448 & 0.6259 & 3.0613 & -0.7876 & 0.9047 & 139.5 & 53.5 \\
1.25 & 0.448 & 0.6259 & 3.0613 & -0.7576 & 0.9047 & 139.8 & 53.7 \\
1.50 & 0.448 & 0.6259 & 3.0613 & -0.7219 & 0.9047 & 140.1 & 53.8 \\
1.75 & 0.448 & 0.6259 & 3.0613 & -0.6812 & 0.9047 & 140.6 & 54.0 \\
2.00 & 0.448 & 0.6259 & 3.0613 & -0.6361 & 0.9047 & 141.4 & 54.3 \\
\end{tabular}
\end{ruledtabular}
\end{table}

\begin{figure}[!htb]
\centering
\includegraphics[scale=0.40]{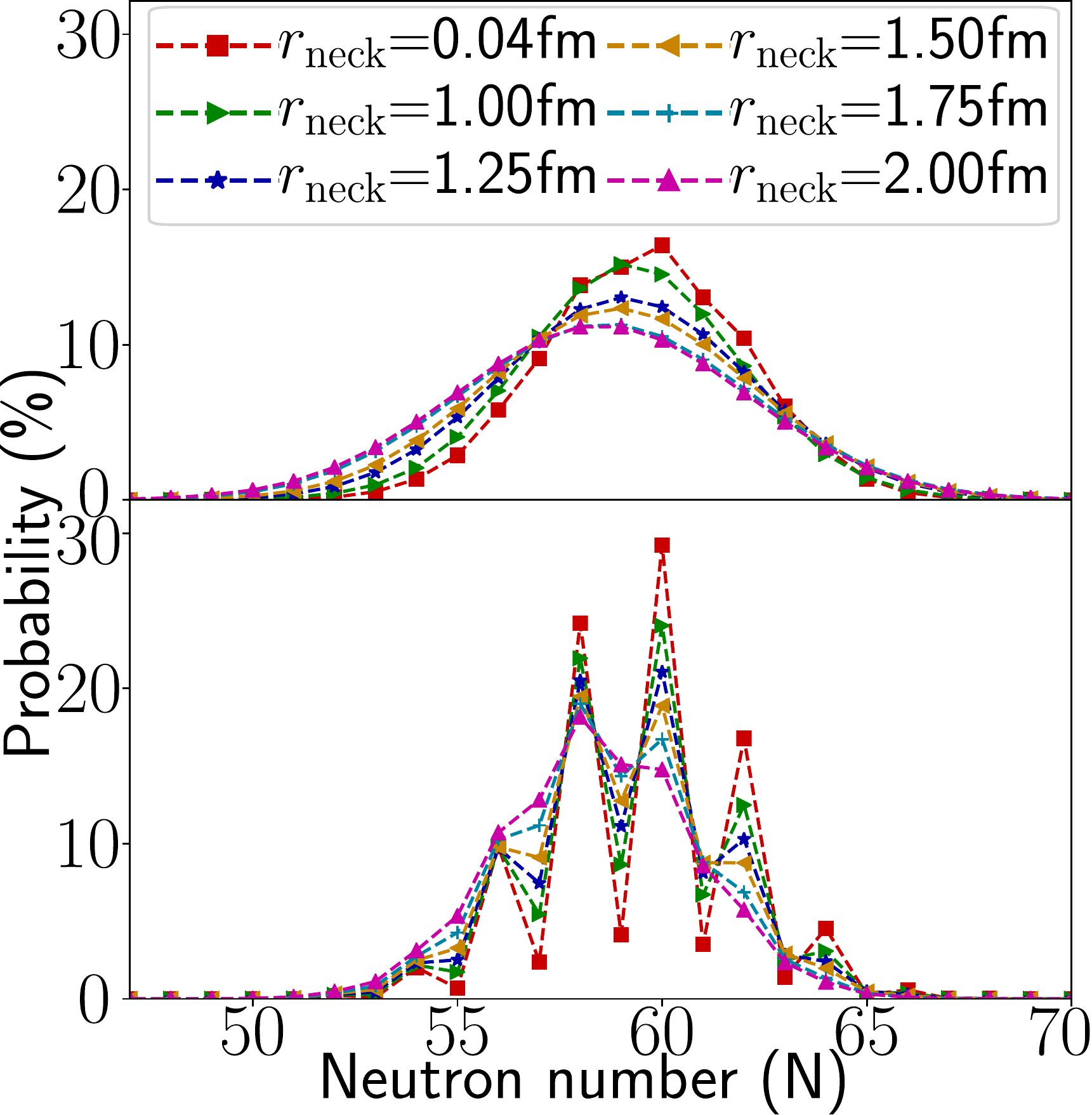}
\caption{Upper panel: Neutron fragmentation probabilities (light fragment) for all the 
configurations listed in Table~\ref{tab:results:neck_shape} obtained with our method. 
Lower panel: same probabilities obtained with the method presented in~\cite{simenel2010}.}
\label{fig:res:YN_neck}
\end{figure}

\begin{figure}[!htb]
\centering
\includegraphics[scale=0.40]{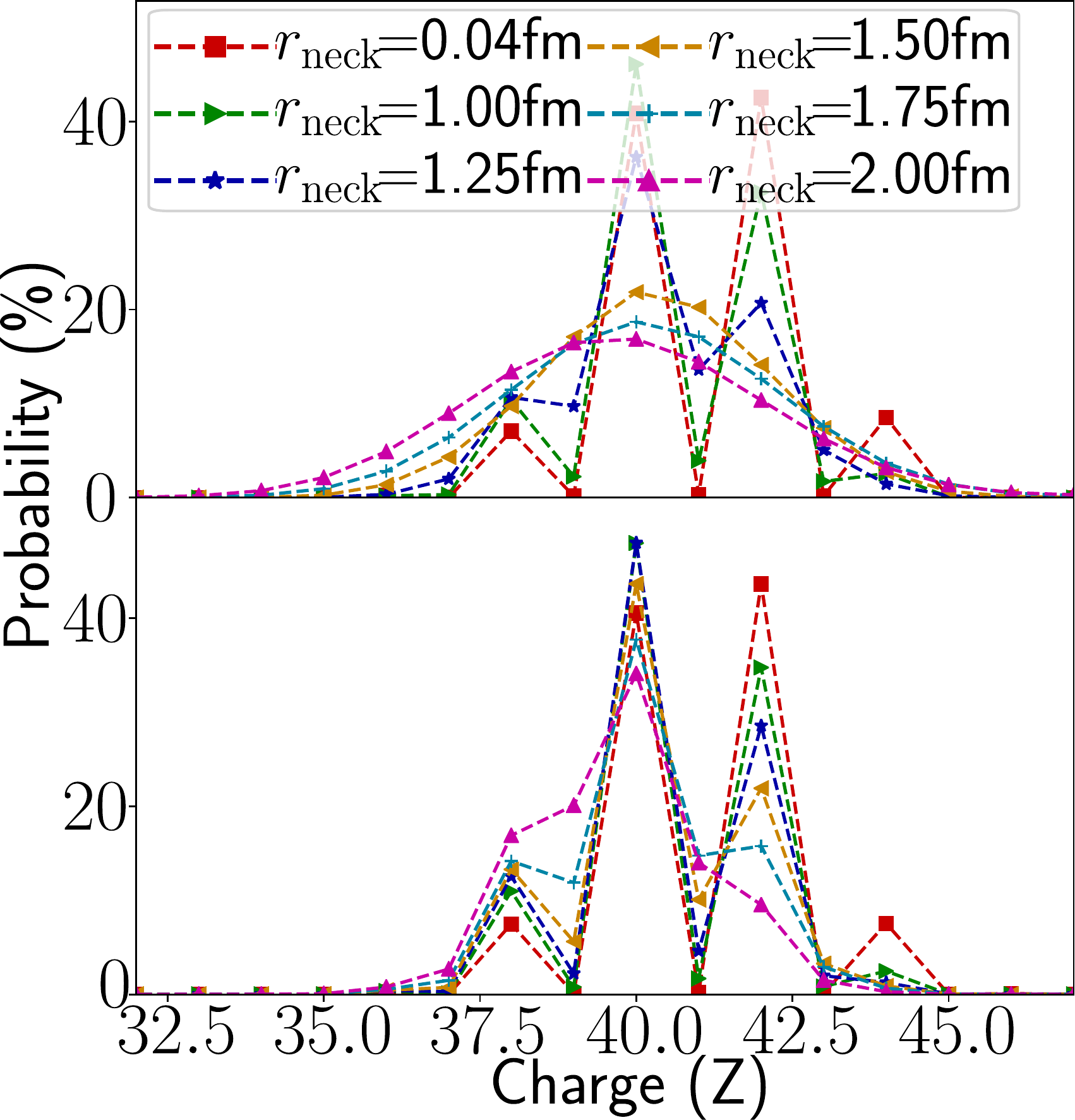}
\caption{Same as \figref{fig:res:YN_neck} for the charge fragmentation probabilities (color 
online).}
\label{fig:res:YZ_neck}
\end{figure}

The neutron fragmentation probabilities are shown in~\figref{fig:res:YN_neck}: the top panel 
shows results with MC sampling, while the bottom panel shows results with projectors. We first 
note that there are substantial differences between the two approaches. Projectors yield an 
OES in the neutron fragmentations probabilities for $r_{\rm neck} < 2.00$ fm. In contrast, 
there is no such OES in the sampling method except for very small values of the neck radius 
($r_{\rm neck} < 0.50$ fm). 

The main explanation for this discrepancy is related to the parametrization of the nuclear 
surface. In generating the family of shapes of \figref{fig:res:shapes_neck}, we started from the 3QS 
parameters corresponding to $r_{\rm neck}=2.5$ fm, and reduced \emph{only} the value of $\sigma_2$ 
to obtain the other shapes. As a consequence of volume conservation, the fragments become 
more and more oblate with decreasing values of $r_{\rm neck}$, which increases the value of 
the neutron Fermi energy. This facilitates tunneling for the states with energies around and above 
the Fermi level, even more so since pairing correlations for neutrons are rather high with 
our parametrization of the pairing force (the pairing gap is of the order of 
$\Delta_{n} \approx 2.25$ MeV for all configurations). This spurious ``geometrical'' effect          
could have been avoided with a more rigorous exploration of the nuclear shapes around our            
initial scission configuration, by making sure that, as we vary the size of the neck, all            
deformation parameters are adjusted so that the the energy remains a minimum. Such an                
exploration of the deformation space is automatic in self-consistent calculations, but               
should be done by hand in semi-phemenological methods. The cost of doing so in a a five-dimensional  
PES as the one we were working with is rather substantial.                                           

Because of the Coulomb barrier, the proton Fermi energy is much lower than the top of the 
barrier between the two prefragments: Tunneling is much less of an issue, and protons are 
better localized. This could explain why the agreement between the two methods is much better 
for the charge fragmentation probabilities, as shown in~\figref{fig:res:YZ_neck}: The 
probability associated with an odd $Z$ vanishes for a value of $r_{\rm neck}$ below 1.00 fm 
with both methods, and a strong OES appears, at $r_{\rm neck} = 1.25$ fm for 
the MC sampling and at $r_{\rm neck} = 1.75$ with projectors. The quantitative agreement 
between the two approaches is, in fact, relatively good for $r_{\rm neck} \leq 1$ fm.

\subsection{EDF Calculations}

To gain further insight, we performed similar calculations in a fully microscopic framework. 
Specifically, we considered the scission configurations near the most likely fission of $^{240}$Pu 
which are discussed extensively in Sec. IV of Ref.~\cite{schunck2014}. These configurations were 
obtained by performing constrained HFB calculations with $\langle \hat{Q}_{20}\rangle = 345$ b 
and $\langle \hat{Q}_{\rm neck} \rangle$ varying between 0.1 and 4.5. All calculations were 
performed with the Skyrme SkM* energy functional; numerical details, such as the size and 
characteristics of the basis, the parameters of the pairing force, etc., can be found in 
Ref.~\cite{schunck2014}. 

\begin{figure}[!htb]
\centering
\includegraphics[scale=0.33]{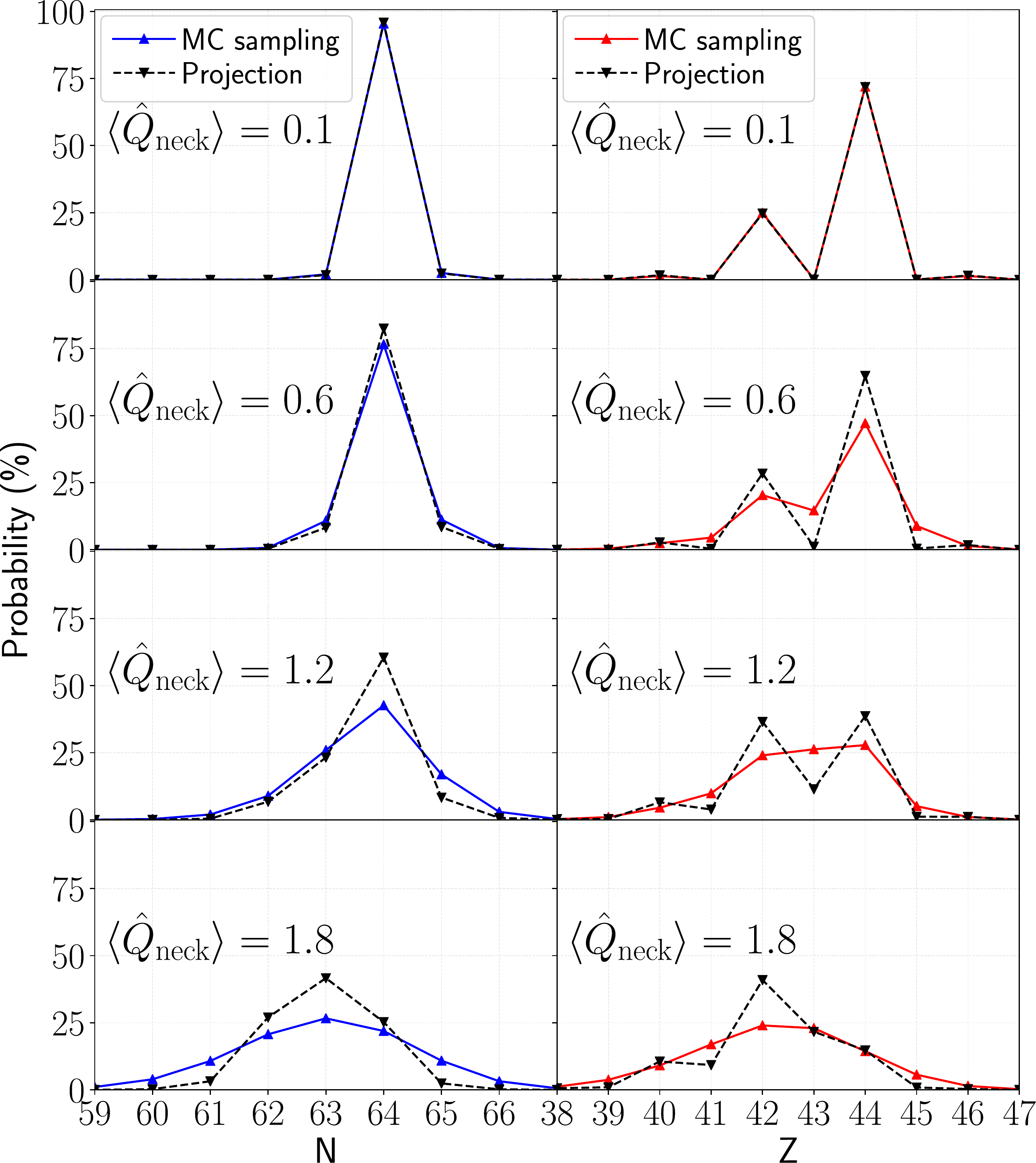}
\caption{Left panel: Comparison of the neutron fragmentation probabilities obtained with our 
Monte Carlo sampling (full lines) and the projection method (dashed black lines) for 
different values of $\langle \hat{Q}_{\rm neck} \rangle$. Right panel: same for the proton 
fragmentations (color online).}
\label{fig:res:YZ_YA_HFB_cmp}
\end{figure}

For each value of the Gaussian neck parameter, we used the double-projection method to 
estimate the fragmentation probabilities of the system. We also computed the occupation 
probabilities $v_{k}^{2}$ in the canonical basis as well as the coefficients $\alpha_{k}$ of 
Eq.~\eqref{eq:ideal:decompo1} needed for the MC method. Fig.~\ref{fig:res:YZ_YA_HFB_cmp} 
summarizes the results for the proton and neutron fragmentation probabilities obtained in 
the two methods. 

We note that the agreement between projection and MC sampling at the limit of very small necks 
is much better than in the macroscopic-microscopic approach. One might attribute this better 
agreement to the fact that, in HFB calculations, the shapes of both the fissioning nucleus and 
the prefragments are automatically determined so that the energy of the fissioning system is 
minimum. As a consequence, such self-consistent calculations do not suffer from the geometrical 
artifact described in the previous section and the Fermi energy of both protons and neutrons 
does not vary much in the range of $\langle \hat{Q}_{\rm neck} \rangle$ considered here, in 
contrast to the macroscopic-microscopic case. 


\section{Conclusion}

We have presented a new method to estimate the uncertainty of particle number in the fission 
fragments. It relies on sampling the probability distribution of finding $N$ particles in the 
fragments based solely on the knowledge of a relevant s.p. basis for the fissioning 
nucleus together with occupation probabilities. We showed that our approach can be used to 
emulate results of particle-number projection techniques, but also provides s.p. 
bases for each subsystem -- provided the latter are sufficiently well separated. We emphasize 
that it is applicable both for Slater determinants and for generalized Slater determinants 
(= quasiparticle vacuum of the HFB theory) and is readily applicable when the energy
states are not degenerate (e.g., when parity is internally broken). Indeed, when parity is 
conserved, s.p. or quasiparticles are, by definition, spread over the two 
prefragments, and the splitting of the individual particle states might lead
to nonorthogonal bases for each parition. In such cases, the degree of orthogonality of
the basis within each partition should be tested to know if the Monte Carlo sampling method
is applicable.

We showed that restoring the particle number in the prefragments formed at scission produces an 
odd-even staggering of probability fragmentations. When combined with full simulations of 
fission dynamics, this result could be key to reproducing the experimentally observed OES 
of the charge distributions. In addition, restoring the particle number could also be 
used to eliminate one of the free parameters typically associated with the calculations of 
fission fragment distributions (folding with a Gaussian, see \cite{regnier2018}). 

While we have illustrated our method in the case of the fission process of heavy atomic 
nuclei, it is, in principle, applicable to a much broader range of problems, such as, for 
example, the localization of electrons inside a molecule. In this case, space partitions 
would correspond to a small volume near each nucleus of the molecule, and we could calculate 
the number of electrons around each of them.


\begin{acknowledgments}
Discussions with Ionel Stetcu, Matt Mumpower and Walid Younes are warmly acknowledged.
This work was performed at Los Alamos National Laboratory, under the auspices of the
National Nuclear Security Administration of the U.S. Department of Energy at Los Alamos
National Laboratory under Contract No. 89233218CNA000001.
Support for this work was provided through the Fission In R-process Elements 
(FIRE) Topical Collaboration in Nuclear Theory of the US Department of Energy. 
It was partly performed under the auspices of the US Department of Energy by 
the Lawrence Livermore National Laboratory under Contract DE-AC52-07NA27344. 
Computing support for this work came from the Lawrence Livermore National 
Laboratory (LLNL) Institutional Computing Grand Challenge program.
\end{acknowledgments} 

\bibliography{bibfile,books}

\end{document}